\documentclass[reprint,amsmath,amssymb,aps]{revtex4-1}

\usepackage{graphicx}% Include figure files
\usepackage{dcolumn}% Align table columns on decimal point
\usepackage{bm}% bold math

\begin{document}

\title{The physics of an optimal basketball free throw}

\author{Irina Barzykina}
\email{irina.barzykina@southbank.net}
\affiliation{Southbank International School, 63-65 Portland Place, London W1B 1QR, UK}

\date{\today}

\begin{abstract}
A physical model is developed, which suggests a pathway to determining the optimal release conditions for a basketball free throw.
Theoretical framework is supported by Monte Carlo simulations and a series of free throws performed and analysed at Southbank International School.
The model defines a smile-shaped success region in angle-velocity space where a free throw will score.
A formula for the minimum throwing angle is derived analytically.
The optimal throwing conditions are determined numerically by maximizing the expected number of successful shots given the error pattern inherent to the player.
Some might need more space for error in velocity, and thus need a higher throwing angle, while others might aim lower because their velocity control is much stronger.
This approach is fully quantified by the model presented and suggests a reliable way for individual free throw improvement. 
The model also explains recent NBA data showing that some of the most successful free throwers bear completely different conditions to the average player.
\end{abstract}

\maketitle

%\tableofcontents

\section{Introduction}

``Does a knowledge of physics help to improve one's basketball skills?'' -- a question asked by Peter Brancazio some 35 years ago \cite{brancazio1981} still presents a considerable challenge and attracts active interest of current researchers \cite{eddings1996,hamilton1997,gablonsky2005,fontanella2006,tran2008,okubo2006,okubo2010,tang2013,schwark2004,silverberg2011,cruzgarza2014,beuoy2015,patankar2016}. 
Physicists usually start their analysis from the free throws. 
Unlike other kinds of throw, it is pretty easy to isolate the variables, since there is no interference from other players. 
Nevertheless, despite all the modern knowledge, professional players only make 75\% of free throws. 
Many players show even less favorable statistics, such as Shaquille O'Neal, notorious for his poor free throw percentage of around 50\% \cite{gablonsky2005,patankar2016}. 
More recent NBA examples are Andre Drummond and DeAndre Jordan hovering around 40\% mark \cite{beuoy2015,patankar2016}.

Although it is hard to expect that a basketball player who steps up to shoot a free throw would think about optimal ball release conditions, knowing those is very important for training. 
Researchers carefully calibrated the forces acting on a basketball in flight to be used in models and simulations \cite{okubo2006,okubo2010,tang2013}. 
Different models have been proposed for optimal shooting trajectory, with subtle differences between them. Broncazio \cite{brancazio1981} emphasized the importance of throwing the ball with the minimum speed, thus allowing for a softer shot. 
Fontanella \cite{fontanella2006} argued that a steeper trajectory makes it easier to score a clean shot, which is what shooters usually aim for. 
Gablonsky and Lang \cite{gablonsky2005} focused on finding the trajectory with the highest margin for error in release angle. 
They calculated the error as a distance from the theoretical boundaries of allowed velocities at a given angle. 

Tran and Silverberg \cite{tran2008} conducted a 3D simulation study to determine once and for all, the physics behind the optimal free throw and later extended their results to bank shots (reflected from backboard) \cite{silverberg2011}. 
Their results allowed them to establish a few guidelines for the foul line: aim toward the back of the rim with 3 Hz of backspin and at $52^{\circ}$ to the horizontal, so that the ball at it’s highest point reaches the top of the backboard. 
However, analysis of NBA’s SportVU motion tracking data showed \cite{beuoy2015} that while a typical $6'3$ (190 cm) player shoots at $54.6^{\circ}$ on average, one of the last year’s  best free shooters of this height, Stephen Curry, was shooting at $58.1^{\circ}$. 
Also James Harden, who became the $11^{\rm{th}}$ in NBA history to make 700 free throws in a season (2014--2015), was shooting at $49.6^{\circ}$ in contrast to NBA average of $53.4^{\circ}$ for his height of $6'5$ (196 cm).
So, not only the average release angle is somewhat higher in professional basketball than physicists suggested, but individual results can differ significantly from the average.
This difference can hardly be attributed to psychological factors alone.
In other words, despite a lot of progress in basketball physics, the optimal throwing conditions are still not fully understood.

This study aims to contribute to resolving this puzzle.
By combining theoretical foundation based on Newton's mechanics, experimental data collected at Southbank International School and the results of Monte Carlo simulations, it will be shown that the optimal throwing conditions are individual and should be considered as such.
In practice, the probability distribution of release parameters of a player should be carefully measured and optimal release conditions determined from simulations on the basis of the determined error statistics. 
Recommendations for individual improvement can then be provided and individual progress monitored with further recursive measurements and recommendations.
This conclusion is in agreement with recent findings by Min \cite{min2016} who suggested to use the phase space volume as the criterion to optimize the shooting strategy.

\section{Theory}

\begin{figure}[tbh!]
 \centering
 \includegraphics[width=8cm]{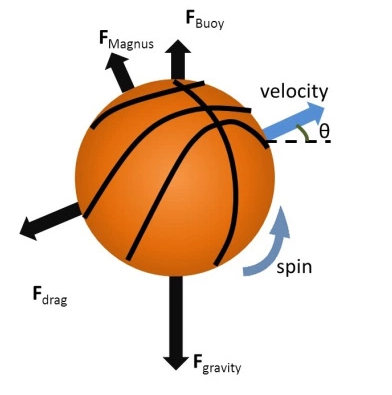}
 \caption{\label{forces}
  The four forces that act on a basketball in flight.
  Reproduced from Ref.~\onlinecite{cruzgarza2014b}.}
\end{figure}

There are four main forces acting upon a basketball in flight. 
They are, in the order of importance: gravity, buoyancy, air resistance and Magnus force \cite{beuoy2015,cruzgarza2014b}, as illustrated in FIG.~\ref{forces}.
The gravitational force pulls the ball vertically down towards the Earth with the corresponding gravitational acceleration $g = 9.81 \rm{m}/\rm{s}^2$.
The buoyant force acts in the opposite direction.
It is caused by the air pressure difference above and below the ball, and its magnitude is proportional to the weight of the air displaced by the ball.
Buoyancy offsets gravity by about $1.5\%$ \cite{beuoy2015}.
Both gravity and buoyancy act on the ball irrespective of its movement.

Air resistance and Magnus force are related to the ball moving through the air.
As the ball moves, it pushes through the air molecules, and the air molecules push back according to Newton's Third Law.
The resulting air resistance force, also known as drag, acts in the opposite direction of the ball's motion; it depends on the ball's velocity, contact area as well as on the density of the air.
Okubo and Hubbard conducted accurate measurements of the air drag force \cite{okubo2010}.
When the ball spins while flying through the air, it experiences uneven friction which creates an unbalanced force perpendicular to the direction of translational motion, called the Magnus force.
Uneven friction is the result of different local speed of the ball surface against the air.
On the side of the ball moving against the air flow a higher pressure zone is created, while on the opposite side of the ball moving with the air flow a lower pressure zone is created.
Most basketball players add backspin to their shot, as a ball with backspin loses more energy on its bounce, which makes it more likely to bounce into the basket \cite{fontanella2006}.
Backspin creates a slight upward lift on the ball's trajectory.

According to Beuoy \cite{beuoy2015} the ball's trajectory can be modeled fairly accurately by ignoring the last two forces, and focusing solely on the combination of gravity and buoyancy.
This combination can be thought of as the effective gravity.
In what follows effective gravity will be implied whenever gravity is mentioned unless specifically stated otherwise.

\subsection{Ball trajectory}

\begin{figure}[tbh!]
 \centering
 \includegraphics[width=8.5cm]{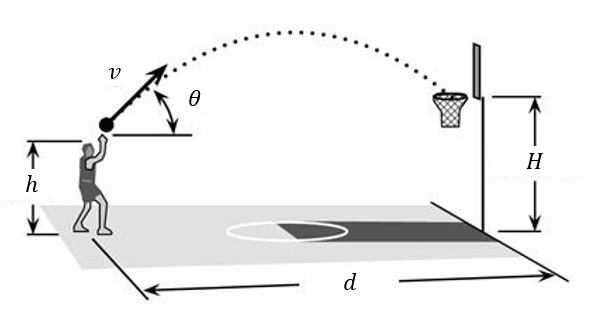}
 \caption{\label{notations}
  Basketball free throw and the corresponding notations: $h$ - the height of ball release, $v$ - initial velocity, $\theta$ - throwing angle, $d$ - horizontal distance from the point of ball release to the center of the hoop, $H$ - height of the hoop.
  }
\end{figure}

When a ball is thrown at a certain angle to the ground its trajectory in-flight is parabolic, unless it is thrown vertically up or down.
Parabolic shape is a consequence of Newton’s laws. 
Let us assume that the ball is released at height $h$ above the ground at angle $\theta$ and velocity $v$, as shown in FIG.~\ref{notations}. 
According to Newton’s First Law, the horizontal velocity component $v \cos \theta$ does not change as there is no horizontal force, so we can write for the $x$ coordinate of the ball at time $t$
\begin{equation}
\label{horizontal}
x = v t \cos \theta
\end{equation}
Here it is assumed that $x=0$ when the ball is released (i.e. when $t=0$). 
Vertical movement occurs under gravity with the gravitational acceleration $g$ pulling the ball to the ground, so as a consequence of Newton’s Second Law the $y$ coordinate at time $t$ should be written as
\begin{equation}
\label{vertical}
y = h + v t \sin \theta -\frac{1}{2} g t^2    
\end{equation}

Here $v \sin \theta$ is the vertical component of the initial velocity and it is assumed that $y(0)=h$. 
By expressing $t$ from Eq.~\ref{horizontal} and substituting into Eq.~\ref{vertical} we find
\begin{equation}
\label{parabola}
y = -x^2\frac{g}{2 v^2 \cos^2 \thetaθ} + x \tan \theta + h
\end{equation}
Using Eq.~\ref{parabola} one can derive a velocity-angle relationship given the target point $(x,y)$.
Since the ball is aiming at the basket, the height of the target is always $y=H$,
\begin{equation}
\label{velocity}
v = \frac{x}{\cos \theta} \sqrt{\frac{g/2}{x \tan \theta + h - H}}
\end{equation}

\subsection{Minimum angle}

\begin{figure}[tbh!]
 \centering
 \includegraphics[width=8.5cm]{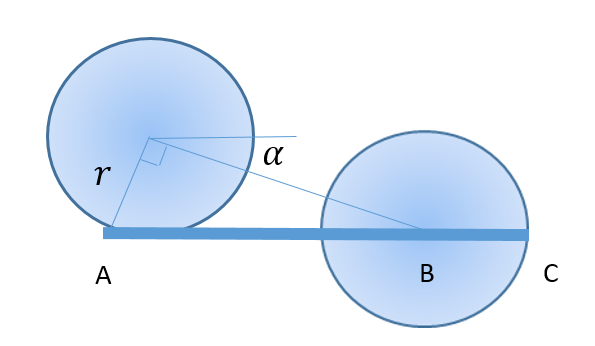}
 \caption{\label{minangle}
  Minimum angle of descent.
  }
\end{figure}

It is clear that in order to score, there should be a minimum angle of release.
The maximum throwing angle is $90^{\circ}$, in the limit.
Let $r$ and $R$ denote the ball and the basket rim radii, respectively. 
The farthest possible point where the ball centre can cross the horizontal rim line is at distance  $x=d+R-r$. 
Otherwise, the back of the rim will be hit and the ball may bounce away from the target.
Now let us assume that the ball descends at angle $\alpha$ and neglect any change in this angle while the ball is above the basket. 
The minimum angle at which the ball can descend without hitting the front of the rim can be found from a rectangular triangle in FIG.~\ref{minangle}, which shows the extreme situation where the ball just passes above the front of the rim and lands just before the back of the rim. 
Given $|AB|+|BC|=|AC|$ it can be shown that
\begin{equation}
\label{sinalpha}
\sin \alpha = \frac{r}{2R-r}
\end{equation}

In order to relate the angle of descent $\alpha$ to the minimum angle of throwing $\theta_{\rm min}$ it can be noticed that $\tan \alpha = v_y / v_x$ or
\begin{equation}
\label{tanalpha}
\tan\alpha = \frac{gt-v\sin\theta_{\rm min}}{v\cos\theta_{\rm min}}
           = \frac{gd}{v^2\cos^2\theta_{\rm min}} - \tan\theta_{\rm min}
\end{equation}
Using Eq.~\ref{velocity} with $x=d$ to exclude velocity from Eq.~\ref{tanalpha} it can be obtained
\begin{equation}
\tan \theta_{\rm min} = \tan \alpha + \frac{2(H-h)}{d}
\end{equation}
Finally, using the identity $\tan^2 \alpha = \sin^2 \alpha / (1-\sin^2 \alpha)$ and Eq.~\ref{sinalpha} it can be shown that
\begin{equation}
\label{thetamin}
\tan \theta_{\rm min} = \frac{r}{2R} \left( 1-\frac{r}{R} \right)^{-1/2} + \frac{2(H-h)}{d}
\end{equation}

\begin{figure}[tbh!]
 \centering
 \includegraphics[width=8.5cm]{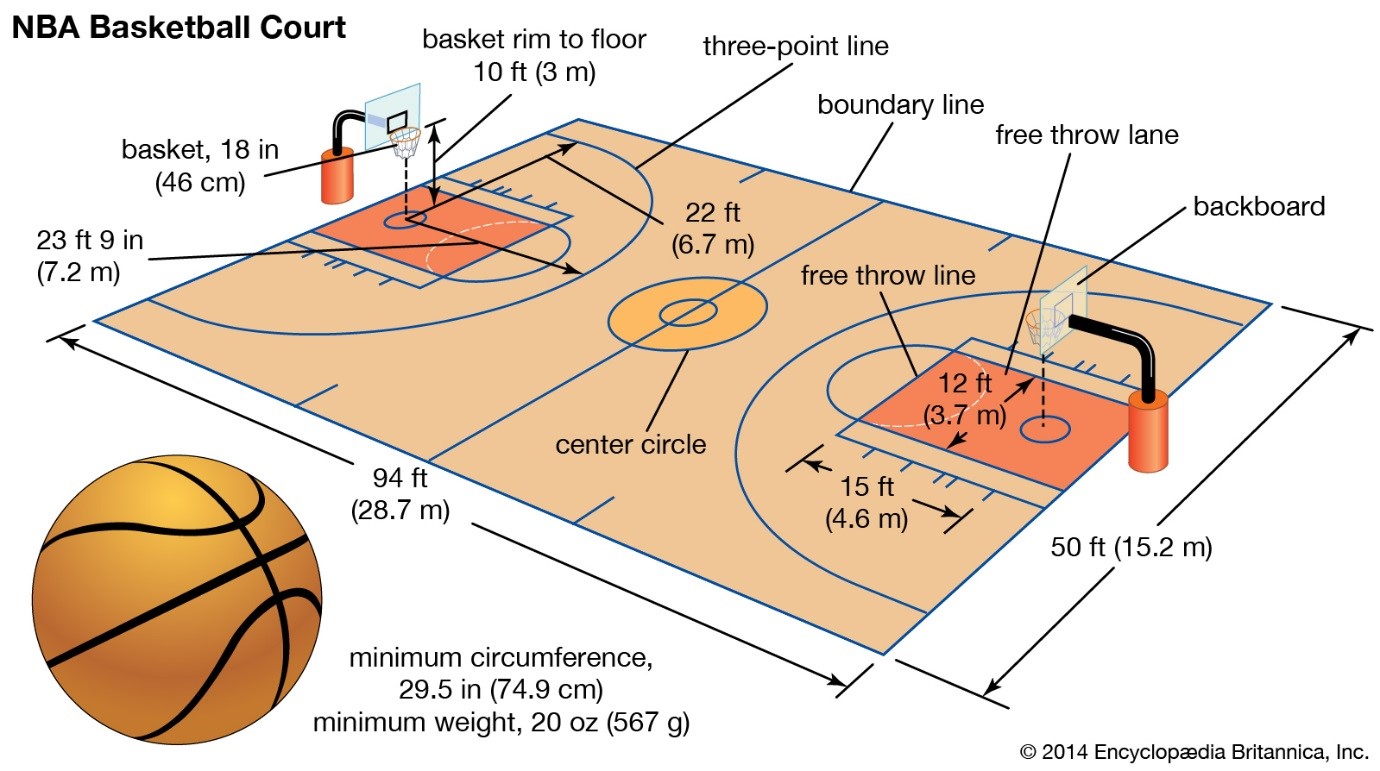}
 \caption{\label{NBACourt}
  NBA basketball court and ball dimensions.
  Reproduced from Ref.~\onlinecite{britannica2014}.
  }
\end{figure}

Eq.~\ref{thetamin} is very important. 
It tells a player of height $h$ (or rather releasing the ball at height $h$) standing a distance $d$ from the basket at which minimum angle to throw. 
The taller the player and the farther he is away from the basket the smaller is the minimum angle. Note that this equation is approximate. 
It was assumed that the trajectory of the ball centre above the basket is almost a straight line. In addition, the relationship between the throwing angle and the angle of descent was calculated by assuming $d \gg R-r$.
These are reasonable assumptions given standard parameters of free throw in basketball, as illustrated in FIG.~\ref{NBACourt}.
These parameters will be used throughout the rest of the paper, namely: $r = 0.12$ m, $R=0.23$ m, $H=3.05$ m, $d=4.6$ m.

\subsection{Angle-velocity `smile'}

When the throwing angle is above the minimum, there is a room for error for the player, i.e. there is a range of allowed velocities for which the target will be hit. 
The maximum velocity is determined when the centre of the ball crosses the horizontal rim line at a distance $d+R-r$, as discussed. 
The minimum velocity is determined when the ball just passes above the front of the rim.
This problem is approached numerically using MATLAB software \cite{matlab2016}.
For each angle $\theta \ge \theta_{\rm min}$ the numerical procedure starts with maximum velocity and decreases it gradually (in small steps of 0.001 m/s) while checking whether the ball hits the front of the rim. 
The trajectory of the ball centre is calculated using Eqs.~\ref{horizontal} and \ref{vertical}, and the condition for not hitting the front of the rim is given by
\begin{equation}
\label{hitfront}
\left(x - (d-R) \right)^2 + \left( y-H \right)^2 > r^2
\end{equation}
This procedure is repeated for a range of angles in steps down to the minimum angle, which is determined numerically by finding the last angle where the trajectory for maximum velocity still satisfies Eq.~\ref{hitfront}.

\begin{figure}[tbh!]
 \centering
 \includegraphics[width=8.5cm]{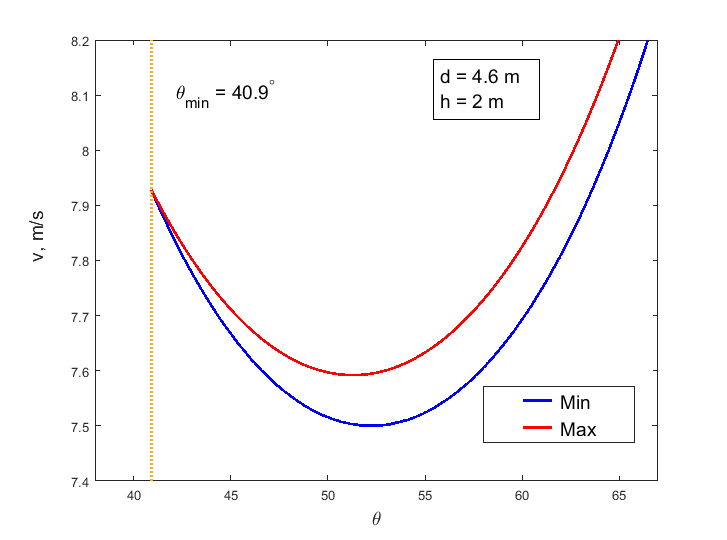}
 \caption{\label{smile}
  The range of angles and velocities that result in a successful free throw for a ball release height of $h=2$ m.
  }
\end{figure}

The results are shown in FIG.~\ref{smile} for the release height of $h=2$ m.
There is a smile-shaped region of release conditions $(\theta,v)$ that result in a successful free throw.
The minimum throwing angle is achieved where the maximum and minimum velocities coincide.
Eq.~\ref{thetamin} predicts $\theta_{\rm min} = 39.8^{\circ}$, which is close but still somewhat underestimates the numerically obtained value of $40.9^{\circ}$.
The minimum angle becomes smaller when the throwing distance becomes larger (not shown).
An interesting observation is that there is a range of angles where the throwing velocity is minimal. 
At this minimal velocity there is a significant room for error for a player. 
For example, if a player throws at 7.59 m/s, he can throw at angles from 47 to $57.5^{\circ}$, nearly 20\% range. 
At the same time, smaller velocity implies better physical control.
This is exactly the result emphasized by Brancazio \cite{brancazio1981}.

\section{Monte Carlo simulations}

\begin{figure}[tbh!]
 \centering
 \includegraphics[width=8.5cm]{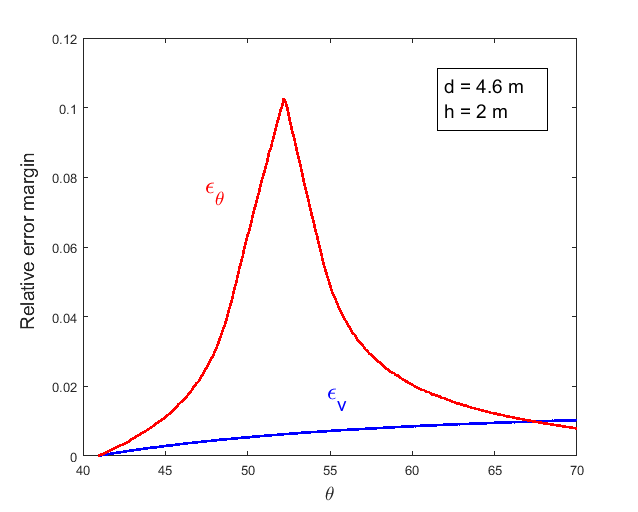}
 \caption{\label{error}
  Relative error margin in angle and velocity as a function of release angle.
  }
\end{figure}

There is always room for error at the moment the ball is released: the error in throwing angle, initial velocity, release height and spin.
This study is focused on angle and velocity, as spin has only little impact on the trajectory and release height is usually stable.
The margin for error can be defined as the minimum distance to the boundary of the `smile' region.
The allowed error in release angle is determined by horizontal (angular) distance to the boundaries, while the error in velocity is determined by vertical distance.
The margin for error is maximized when the distance to both boundaries is equal.
FIG.~\ref{error} shows the relative error margin in angle and velocity.
While the allowed error in release angle is maximized in the area of minimum velocity, this does not mean that this area is optimal overall.
The allowed error in velocity increases with increasing angle.
So if the shooter was perfect in angle, he would have to shoot at higher angles to maximize the expected error in velocity.
This idea was discussed by Fontanella \cite{fontanella2006}.
Gablonsky and Lang \cite{gablonsky2005} noticed that in order to define optimal release conditions one has to maximize the expected error in both angle and velocity.
They considered this problem from the viewpoint of multiobjective optimization using a heuristic argument that the best trajectory is the one that puts five times as much emphasis on the error in velocity than on the error in angle.

\begin{figure}[tbh!]
 \centering
 \includegraphics[width=8.5cm]{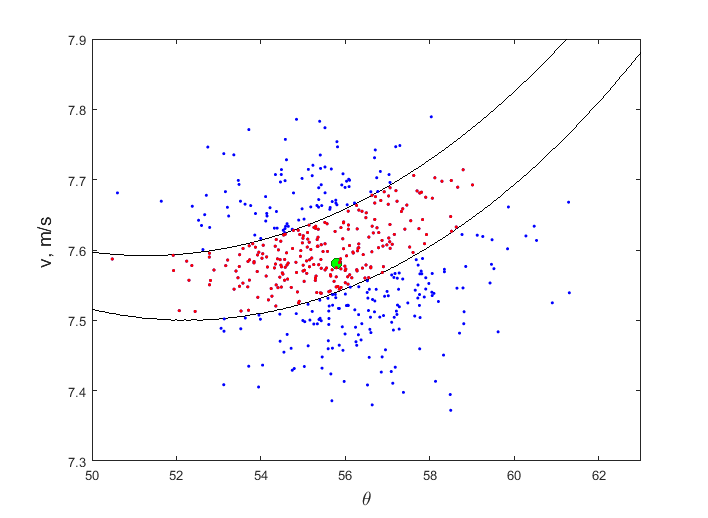}
 \caption{\label{sim}
  Monte Carlo simulation procedure. 
  Green circle is the target point. 
  Dots are random realizations. 
  Black lines are success bounds.
  Red dots are scored, blue dots are missed.
  }
\end{figure}

The approach taken in this paper is different.
It is suggested to start from the probability distribution of angle and velocity, that is, when the ball is released with certain target values of $\theta_0$ and $v_0$ the realized values can be different but governed by the probability distribution centered around $(\theta_0, v_0)$.
It is assumed for simplicity that errors in angle and velocity are independent and Gaussian, and characterized by standard deviations $\sigma_{\theta}$ and $\sigma_v$, respectively.
Then in a Monte Carlo simulation, a million of different release conditions are generated for a given target pair $(\theta_0, v_0)$, using MATLAB built-in normal random number generator.
It is checked whether realized $(\theta, v)$ values would result in successful shots (would hit the `smile' zone) and the total probability of success is counted as a fraction of successful realizations, as illustrated in FIG.~\ref{sim}.
\begin{figure}[tbh!]
 \centering
 \includegraphics[width=8.5cm]{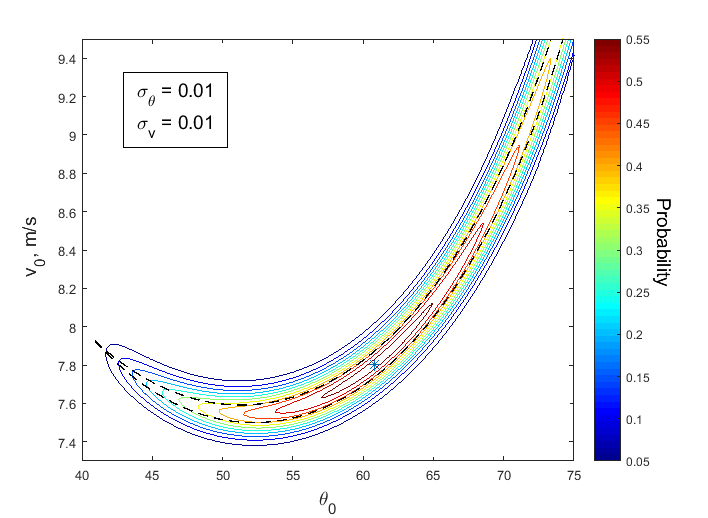}
 \caption{\label{contour}
  Contour plot of score probability distribution as a function of target release angle $\theta_0$ and target release velocity $v_0$.
  Errors are assumed to be normal with relative standard deviations $\sigma_{\theta} = \sigma_v = 0.01$.
  Release height was $h=2$ m.
  Dashed black lines are success bounds.
  Star denotes the point of maximum score probability.
  }
\end{figure}

\begin{figure}[tbh!]
 \centering
 \includegraphics[width=8.5cm]{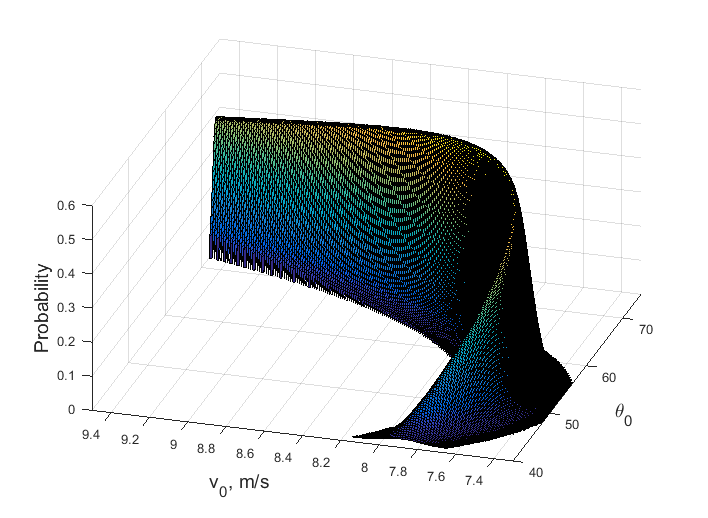}
 \caption{\label{surf}
  Surface plot of score probability distribution corresponding to FIG.~\ref{contour}.
  }
\end{figure}

The procedure is repeated on a rectangular grid of release angle-velocity pairs $(\theta_0, v_0)$, providing score probability for each pair.
The resulting numerical score probability distribution is illustrated in FIG.~\ref{contour} (contour plot) and FIG.~\ref{surf} (surface plot).
It can be seen that when both angle and velocity can be executed with the same error of one percent, optimal release conditions move to higher angle and higher velocity.

\begin{figure}[tbh!]
 \centering
 \includegraphics[width=8.5cm]{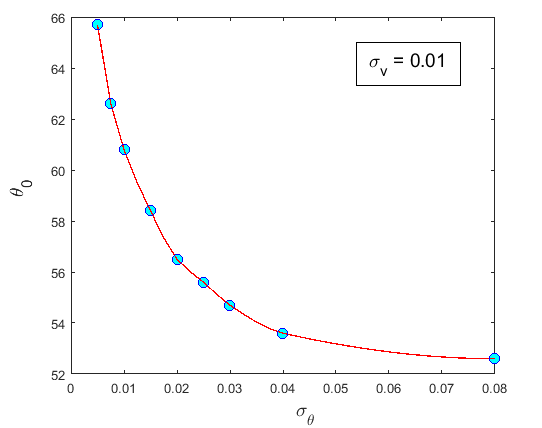}
 \caption{\label{optimaltheta}
  Optimal release angle as a function of $\sigma_{\theta}$.
 }
\end{figure}

\begin{figure}[tbh!]
 \centering
 \includegraphics[width=8.5cm]{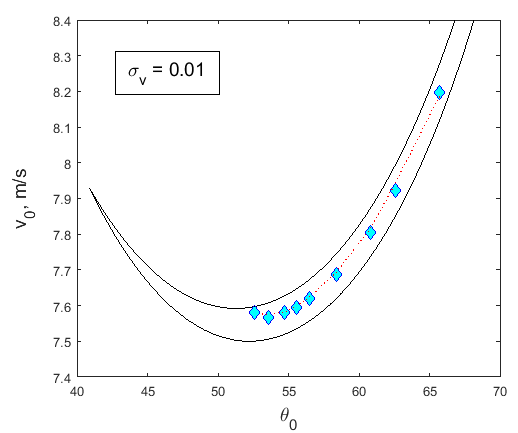}
 \caption{\label{optimalrelease}
  Optimal release angle-velocity pairs (diamonds) as a function of angle percent error, corresponding to FIG.~\ref{optimaltheta}.
  Black lines are success bounds.
 }
\end{figure}

This effect can be investigated further by repeating the whole Monte Carlo simulation procedure for different values of $\sigma_{\theta}$ while keeping $\sigma_v = 0.01$, and finding optimal $(\theta_0, v_0)$ pair in each case.
As it can be seen in FIG.~\ref{optimaltheta} and FIG.~\ref{optimalrelease}, as the relative error in throwing angle increases, the optimal throwing angle decreases and the optimal release point moves towards the minimum velocity area.
This area allows the shooter the maximum error in angle.
In the opposite scenario, as the relative error in throwing angle decreases, the optimal throwing angle increases balancing the error in velocity.
It is impossible to expect that a player can execute the throw with absolute precision in velocity.
Therefore, one should expect to have optimal throwing conditions to be shifted to somewhat higher angles.

\section{Experiment}

\begin{figure}[tbh!]
 \centering
 \includegraphics[width=8.5cm]{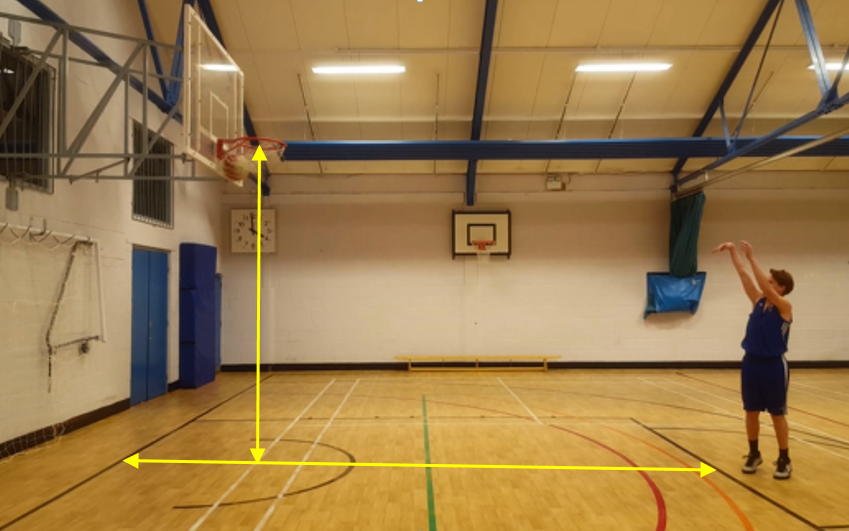}
 \caption{\label{lucameasures}
  Free throw setup and standard measures used to determine the scaling coefficient.
 }
\end{figure}

It order to verify theoretical conclusions, a series of 25 free throws by a student of Southbank International School was recorded on video.
It is certainly not as detailed a study as the one based on NBA data \cite{beuoy2015}.
However, real data will help reinforce qualitative conclusions.
The student scored 52\%, an outstanding result for a non-professional.
Each individual shot was processed frame by frame to determine the height of release $h$, the horizontal distance travelled by the ball $d$, in-flight time $t$ and the maximum height in balls trajectory $h_{\rm max}$.
Time was measured in frames, with each frame being $1/30^{\rm th}$ of a second.
Distance was measured in mm right on the image.
In order to determine scaling coefficient, one standard horizontal distance was measured, the distance to free throw line of 5.79 m, and one standard vertical distance, the height of the hoop, $H=3.05$ m, as shown in FIG.~\ref{lucameasures}.
Both horizontal and vertical scaling was consistent providing a unique scaling coefficient of 19.4.
The difficulty (and additional error) in such measurements comes from 3D projection.
It is obvious in FIG.~\ref{lucameasures} that the horizontal lines at the bottom of the screen are tilted downwards while those at the top are tilted upwards.

The height of release was very stable at $h=2.02$ m in all measurements.
This is player's height with his arms extended.
The error of this figure is definitely more than just a 1\% distance measurement error (defined by the ruler).
Firstly, the ball position during release is somewhat smeared.
Secondly, the player is located on the right of the image where distance is amplified due to 3D projection effect.
The resulting error is estimated to be at least 5\% (knowing the height of the player).
Improvement can be achieved by using a grid of cameras but this is outside of scope of the present study.

The parameters were determined as follows: $v_x = d/t$, $v_y = \sqrt{2g(h_{\rm max}-h)}$, $v = \sqrt{v_x^2 + v_y^2}$, $\theta = \arctan(v_y/v_x)$.
The $g$ factor was taken to be 9.66 to account for buoyancy.
The results are summarized in TABLE~\ref{resultstable}.

\begin{table}[h]
\centering
\begin{tabular}{|c|c|c|c|c|c|c|c|}
\hline
$d$, m & $h_{\rm max}$, m & $t$, s & $v_x$, m/s & $v_y$, m/s & $v$, m/s & $\theta$, $^{\circ}$ & score \\
\hline
4.55 & 4.04	& 0.94 & 4.26 &	6.25 & 7.49	& 55.71	& 1 \\
\hline
4.59 & 4.12	& 0.91 & 4.18 & 6.37 & 7.61 & 56.69 & 1 \\
\hline
4.22 & 4.08 & 0.92 & 3.90 & 6.31 & 7.38 & 58.29 & 0 \\
\hline
4.61 & 4.06 & 0.93 & 4.29 & 6.28 & 7.54 & 55.68 & 1 \\
\hline
4.76 & 4.08 & 0.92 & 4.40 & 6.31 & 7.65 & 55.10 & 0 \\
\hline
4.24 & 4.10 & 0.92 & 3.89 & 6.34 & 7.41 & 58.46 & 0 \\
\hline
4.53 & 4.10 & 0.92 & 4.16 & 6.34 & 7.55 & 56.74 & 1 \\
\hline
4.12 & 3.83 & 1.02 & 4.19 & 5.91 & 6.99 & 54.65 & 0 \\
\hline
4.57 & 4.16 & 0.90 & 4.11 & 6.43 & 7.65 & 57.38 & 1 \\
\hline
4.57 & 4.14 & 0.91 & 4.14 & 6.40 & 7.62 & 57.09 & 1 \\
\hline
4.55 & 4.02 & 0.94 & 4.29 & 6.22 & 7.46 & 55.40 & 1 \\
\hline
4.30 & 3.98 & 0.96 & 4.11 & 6.16 & 7.29 & 56.28 & 0 \\
\hline
4.61 & 4.12 & 0.91 & 4.20 & 6.37 & 7.62 & 56.58 & 1 \\
\hline
4.20 & 3.94 & 0.97 & 4.08 & 6.10 & 7.19 & 56.24 & 0 \\
\hline
4.59 & 4.08 & 0.92 & 4.24 & 6.31 & 7.56 & 56.10 & 1 \\
\hline
4.55 & 4.08 & 0.92 & 4.20 & 6.31 & 7.54 & 56.32 & 1 \\
\hline
4.86 & 4.18 & 0.89 & 4.35 & 6.46 & 7.82 & 56.04 & 0 \\
\hline
4.86 & 4.10 & 0.92 & 4.46 & 6.34 & 7.72 & 54.86 & 0 \\
\hline
4.34 & 4.02 & 0.94 & 4.09 & 6.22 & 7.36 & 56.68 & 0 \\
\hline
4.30 & 3.85 & 1.01 & 4.33 & 5.94 & 7.10 & 53.89 & 0 \\
\hline
4.61 & 4.00 & 0.95 & 4.38 & 6.19 & 7.47 & 54.74 & 1 \\
\hline
4.59 & 4.06 & 0.93 & 4.27 & 6.28 & 7.53 & 55.79 & 1 \\
\hline
4.40 & 3.85 & 1.01 & 4.43 & 5.94 & 7.15 & 53.28 & 0 \\
\hline
4.22 & 3.89 & 0.99 & 4.19 & 6.00 & 7.12 & 55.10 & 0 \\
\hline
4.63 & 4.10 & 0.92 & 4.25 & 6.34 & 7.60 & 56.18 & 1 \\
\hline
\end{tabular}
\caption{\label{resultstable}
    Details of 25 consecutive free throws by a student of Southbank International School.    
}
\end{table}

\begin{figure}[tbh!]
 \centering
 \includegraphics[width=8.5cm]{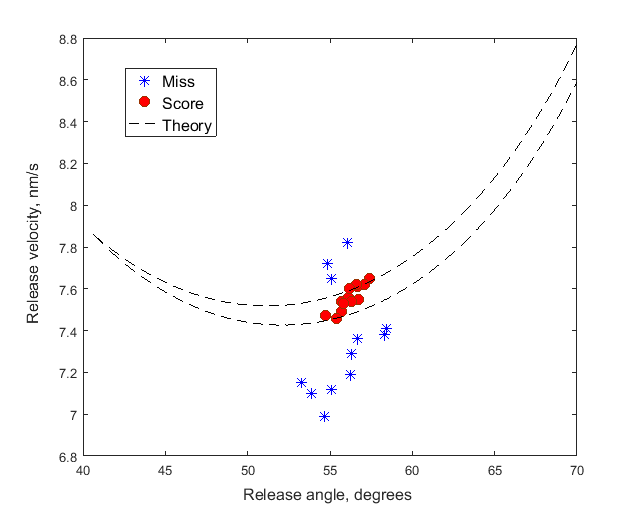}
 \caption{\label{luca}
  Angle-velocity plot of 25 free throws detailed in TABLE~\ref{resultstable}.
  Red circles correspond to successful shots, blue stars show where the target was missed.
  Dashed black lines are theoretical success bounds.
 }
\end{figure}

Experimental results are further compared to the theoretical success bounds in FIG.~\ref{luca}.
Excellent agreement is seen -- successful shots all basically fit into the success zone, despite the error involved in measurements.
It is interesting to notice that the observed relative standard deviations $\sigma_{\theta} = 0.022$ and $\sigma_v = 0.028$ of release angle and velocity, respectively, are quite close for this particular player.
Therefore, according to the results of the previous section, he should be aiming higher than the minimum velocity area.
And indeed he does, with the average release angle being $56^{\circ}$.

\section{Concluding remarks}

A physical model was developed, which suggests a pathway to determining the optimal release conditions for a basketball free throw.
The model was supported by Monte Carlo simulations and a series of free throws performed by a student of Southbank International School.
In agreement with previous findings in the literature, the model confirms that tall players should generally be better shooters as they have more room for error and their required optimal release angle should generally be lower \cite{beuoy2015}.
The model also suggests that the optimal shot should aim somewhere between the center of the basket and the back of the rim \cite{gablonsky2005}.
Back spin was not discussed but it is generally considered useful to secure shots where the ball collides with the rim or the board \cite{fontanella2006, tran2008}.
The model and all the conclusions apply equally well for wheelchair basketball \cite{schwark2004}.

What was particularly emphasized in this study was that the optimal release conditions vary from player to player, not only because they have different release heights, but also because of their different levels of consistency in release angles and velocities.
If one records a series of throws by a player and calculates probability distribution of errors in release angle and velocity, then using the technique demonstrated in this paper it would be possible to determine optimal throwing conditions that would maximize the expected number of successful shots given the error pattern inherent to the player.
Recommendations for individual improvement can then be provided and individual progress monitored with further recursive measurements and recommendations.
This approach can be extended to more realistic Monte Carlo simulation including all the relevant forces \cite{min2016} and empirical probability distribution in angle-velocity space.

Recent analysis of NBA data shows that some of the most successful free throwers bear completely different conditions to the average player \cite{beuoy2015}, which also confirms the fact that all players have different consistency in release angles and velocities.
Some might need more space for error in velocity, and thus need a higher throwing angle, while others might aim lower because their velocity control is much stronger.
The physical model discussed in this paper fully supports this point of view and shows the way to translate the error footprint of a player into individual conditions for optimal free throw.

It is well known that psychological aspects may play an extremely important role in influencing ones scoring ability.
With support of a good model one can confidently practice optimal shots to the degree of automation and thus should be able to keep nerves out of the way.

\begin{acknowledgments}

This work was performed as part of IB Personal Project at Southbank International School under supervision of Mr S. Aylward.
The author is grateful to Mr Aylward for fruitful discussions and critical comments.
Support of PE Department of Southbank International School is highly appreciated.
The author would like to thank Mr S. Hajjaj and Mr E. Clarke for insightful discussions about basketball techniques and psychology.
Special thanks are due to Luca Bastoni who performed a series of free throws.

\end{acknowledgments}

\bibliography{bibliography}

\begin{thebibliography}{18}%
\makeatletter
\providecommand \@ifxundefined [1]{%
 \@ifx{#1\undefined}
}%
\providecommand \@ifnum [1]{%
 \ifnum #1\expandafter \@firstoftwo
 \else \expandafter \@secondoftwo
 \fi
}%
\providecommand \@ifx [1]{%
 \ifx #1\expandafter \@firstoftwo
 \else \expandafter \@secondoftwo
 \fi
}%
\providecommand \natexlab [1]{#1}%
\providecommand \enquote  [1]{``#1''}%
\providecommand \bibnamefont  [1]{#1}%
\providecommand \bibfnamefont [1]{#1}%
\providecommand \citenamefont [1]{#1}%
\providecommand \href@noop [0]{\@secondoftwo}%
\providecommand \href [0]{\begingroup \@sanitize@url \@href}%
\providecommand \@href[1]{\@@startlink{#1}\@@href}%
\providecommand \@@href[1]{\endgroup#1\@@endlink}%
\providecommand \@sanitize@url [0]{\catcode `\\12\catcode `\$12\catcode
  `\&12\catcode `\#12\catcode `\^12\catcode `\_12\catcode `\%12\relax}%
\providecommand \@@startlink[1]{}%
\providecommand \@@endlink[0]{}%
\providecommand \url  [0]{\begingroup\@sanitize@url \@url }%
\providecommand \@url [1]{\endgroup\@href {#1}{\urlprefix }}%
\providecommand \urlprefix  [0]{URL }%
\providecommand \Eprint [0]{\href }%
\providecommand \doibase [0]{http://dx.doi.org/}%
\providecommand \selectlanguage [0]{\@gobble}%
\providecommand \bibinfo  [0]{\@secondoftwo}%
\providecommand \bibfield  [0]{\@secondoftwo}%
\providecommand \translation [1]{[#1]}%
\providecommand \BibitemOpen [0]{}%
\providecommand \bibitemStop [0]{}%
\providecommand \bibitemNoStop [0]{.\EOS\space}%
\providecommand \EOS [0]{\spacefactor3000\relax}%
\providecommand \BibitemShut  [1]{\csname bibitem#1\endcsname}%
\let\auto@bib@innerbib\@empty
%</preamble>
\bibitem [{\citenamefont {Brancazio}(1981)}]{brancazio1981}%
  \BibitemOpen
  \bibfield  {author} {\bibinfo {author} {\bibfnamefont {P.~J.}\ \bibnamefont
  {Brancazio}},\ }\href {\doibase 10.1119/1.12511} {\bibfield  {journal}
  {\bibinfo  {journal} {Amer.~J.~Phys.}\ }\textbf {\bibinfo {volume} {49}},\
  \bibinfo {pages} {356} (\bibinfo {year} {1981})}\BibitemShut {NoStop}%
\bibitem [{\citenamefont {Eddings}(1996)}]{eddings1996}%
  \BibitemOpen
  \bibfield  {author} {\bibinfo {author} {\bibfnamefont {M.~R.}\ \bibnamefont
  {Eddings}},\ }\href {http://www.csuchico.edu/~jhudson/pdf/eddings_thesis.pdf}
  {\emph {\bibinfo {title} {Effect of manipulating angle of projection on
  height of release and accuracy in the basketball free throw: A biomechanical
  study}}}\ (\bibinfo  {publisher} {Master's thesis, California State
  University},\ \bibinfo {address} {Chico},\ \bibinfo {year}
  {1996})\BibitemShut {NoStop}%
\bibitem [{\citenamefont {Hamilton}\ and\ \citenamefont
  {Reinschmidt}(1997)}]{hamilton1997}%
  \BibitemOpen
  \bibfield  {author} {\bibinfo {author} {\bibfnamefont {G.~R.}\ \bibnamefont
  {Hamilton}}\ and\ \bibinfo {author} {\bibfnamefont {C.}~\bibnamefont
  {Reinschmidt}},\ }\href {\doibase 10.1080/026404197367137} {\bibfield
  {journal} {\bibinfo  {journal} {J.~Sports Sci.}\ }\textbf {\bibinfo {volume}
  {15}},\ \bibinfo {pages} {491} (\bibinfo {year} {1997})}\BibitemShut
  {NoStop}%
\bibitem [{\citenamefont {Gablonsky}\ and\ \citenamefont
  {Lang}(2005)}]{gablonsky2005}%
  \BibitemOpen
  \bibfield  {author} {\bibinfo {author} {\bibfnamefont {J.~M.}\ \bibnamefont
  {Gablonsky}}\ and\ \bibinfo {author} {\bibfnamefont {A.~S. I.~D.}\
  \bibnamefont {Lang}},\ }\href {\doibase 10.1137/S0036144598339555} {\bibfield
   {journal} {\bibinfo  {journal} {SIAM Rev.}\ }\textbf {\bibinfo {volume}
  {47}},\ \bibinfo {pages} {775} (\bibinfo {year} {2005})}\BibitemShut
  {NoStop}%
\bibitem [{\citenamefont {Fontanella}(2006)}]{fontanella2006}%
  \BibitemOpen
  \bibfield  {author} {\bibinfo {author} {\bibfnamefont {J.~J.}\ \bibnamefont
  {Fontanella}},\ }\href@noop {} {\emph {\bibinfo {title} {The physics of
  basketball}}}\ (\bibinfo  {publisher} {The John Hopkins University Press},\
  \bibinfo {address} {Baltimore},\ \bibinfo {year} {2006})\BibitemShut
  {NoStop}%
\bibitem [{\citenamefont {Tran}\ and\ \citenamefont
  {Silverberg}(2008)}]{tran2008}%
  \BibitemOpen
  \bibfield  {author} {\bibinfo {author} {\bibfnamefont {C.~M.}\ \bibnamefont
  {Tran}}\ and\ \bibinfo {author} {\bibfnamefont {L.~M.}\ \bibnamefont
  {Silverberg}},\ }\href {\doibase 10.1080/02640410802004948} {\bibfield
  {journal} {\bibinfo  {journal} {J.~Sports Sci.}\ }\textbf {\bibinfo {volume}
  {26}},\ \bibinfo {pages} {1147} (\bibinfo {year} {2008})}\BibitemShut
  {NoStop}%
\bibitem [{\citenamefont {Okubo}\ and\ \citenamefont
  {Hubbard}(2006)}]{okubo2006}%
  \BibitemOpen
  \bibfield  {author} {\bibinfo {author} {\bibfnamefont {H.}~\bibnamefont
  {Okubo}}\ and\ \bibinfo {author} {\bibfnamefont {M.}~\bibnamefont
  {Hubbard}},\ }\href {\doibase 10.1080/02640410500520401} {\bibfield
  {journal} {\bibinfo  {journal} {J.~Sports Sci.}\ }\textbf {\bibinfo {volume}
  {24}},\ \bibinfo {pages} {1303} (\bibinfo {year} {2006})}\BibitemShut
  {NoStop}%
\bibitem [{\citenamefont {Okubo}\ and\ \citenamefont
  {Hubbard}(2010)}]{okubo2010}%
  \BibitemOpen
  \bibfield  {author} {\bibinfo {author} {\bibfnamefont {H.}~\bibnamefont
  {Okubo}}\ and\ \bibinfo {author} {\bibfnamefont {M.}~\bibnamefont
  {Hubbard}},\ }\href {\doibase 10.1016/j.proeng.2010.04.145} {\bibfield
  {journal} {\bibinfo  {journal} {Procedia Engineering}\ }\textbf {\bibinfo
  {volume} {2}},\ \bibinfo {pages} {3281} (\bibinfo {year} {2010})}\BibitemShut
  {NoStop}%
\bibitem [{\citenamefont {Tang}\ \emph {et~al.}(2013)\citenamefont {Tang},
  \citenamefont {Ma},\ and\ \citenamefont {Guo}}]{tang2013}%
  \BibitemOpen
  \bibfield  {author} {\bibinfo {author} {\bibfnamefont {D.}~\bibnamefont
  {Tang}}, \bibinfo {author} {\bibfnamefont {G.}~\bibnamefont {Ma}}, \ and\
  \bibinfo {author} {\bibfnamefont {J.}~\bibnamefont {Guo}},\ }\href {\doibase
  10.3923/itj.2013.3315.3319} {\bibfield  {journal} {\bibinfo  {journal}
  {Inform.~Technol.~J.}\ }\textbf {\bibinfo {volume} {12}},\ \bibinfo {pages}
  {3315} (\bibinfo {year} {2013})}\BibitemShut {NoStop}%
\bibitem [{\citenamefont {Schwark}\ \emph {et~al.}(2004)\citenamefont
  {Schwark}, \citenamefont {Mackenzie},\ and\ \citenamefont
  {Sprigings}}]{schwark2004}%
  \BibitemOpen
  \bibfield  {author} {\bibinfo {author} {\bibfnamefont {B.~N.}\ \bibnamefont
  {Schwark}}, \bibinfo {author} {\bibfnamefont {S.~J.}\ \bibnamefont
  {Mackenzie}}, \ and\ \bibinfo {author} {\bibfnamefont {E.~J.}\ \bibnamefont
  {Sprigings}},\ }\href {\doibase 10.1123/jab.20.2.153} {\bibfield  {journal}
  {\bibinfo  {journal} {J.~Appl.~Biomech.}\ }\textbf {\bibinfo {volume} {20}},\
  \bibinfo {pages} {153} (\bibinfo {year} {2004})}\BibitemShut {NoStop}%
\bibitem [{\citenamefont {Silverberg}\ \emph {et~al.}(2011)\citenamefont
  {Silverberg}, \citenamefont {Tran},\ and\ \citenamefont
  {Adams}}]{silverberg2011}%
  \BibitemOpen
  \bibfield  {author} {\bibinfo {author} {\bibfnamefont {L.~M.}\ \bibnamefont
  {Silverberg}}, \bibinfo {author} {\bibfnamefont {C.~M.}\ \bibnamefont
  {Tran}}, \ and\ \bibinfo {author} {\bibfnamefont {T.~M.}\ \bibnamefont
  {Adams}},\ }\href {\doibase 10.2202/1559-0410.1299} {\bibfield  {journal}
  {\bibinfo  {journal} {J.~Appl.~Biomech.}\ }\textbf {\bibinfo {volume} {7}}
  (\bibinfo {year} {2011}),\ 10.2202/1559-0410.1299}\BibitemShut {NoStop}%
\bibitem [{\citenamefont {Cruz-Garza}(2014{\natexlab{a}})}]{cruzgarza2014}%
  \BibitemOpen
  \bibfield  {author} {\bibinfo {author} {\bibfnamefont {J.~G.}\ \bibnamefont
  {Cruz-Garza}},\ }\href@noop {} {\enquote {\bibinfo {title} {\emph{Optimal
  trajectory for a shot}},}\ }\bibinfo {howpublished}
  {\url{physicsofbasketball.wordpress.com/2014/05/18/optimal-trajectory-for-a-shot}}
  (\bibinfo {year} {2014}{\natexlab{a}}),\ \bibinfo {note} {online May 18,
  2014}\BibitemShut {NoStop}%
\bibitem [{\citenamefont {Beuoy}(2015)}]{beuoy2015}%
  \BibitemOpen
  \bibfield  {author} {\bibinfo {author} {\bibfnamefont {M.}~\bibnamefont
  {Beuoy}},\ }\href@noop {} {\enquote {\bibinfo {title} {\emph{Introducing
  ShArc: Shot arc analysis}},}\ }\bibinfo {howpublished}
  {\url{www.inpredictable.com/2015/05/introducing-sharc-shot-arc-analysis.html}}
  (\bibinfo {year} {2015}),\ \bibinfo {note} {online May 26, 2015}\BibitemShut
  {NoStop}%
\bibitem [{\citenamefont {Patankar}(2016)}]{patankar2016}%
  \BibitemOpen
  \bibfield  {author} {\bibinfo {author} {\bibfnamefont {V.}~\bibnamefont
  {Patankar}},\ }\href@noop {} {\enquote {\bibinfo {title} {\emph{What a dairy
  farmer can teach Shaq about free throws}},}\ }\bibinfo {howpublished}
  {\url{www.process.st/how-to-shoot-a-free-throw}} (\bibinfo {year} {2016}),\
  \bibinfo {note} {online January 8, 2016}\BibitemShut {NoStop}%
\bibitem [{\citenamefont {Min}(2016)}]{min2016}%
  \BibitemOpen
  \bibfield  {author} {\bibinfo {author} {\bibfnamefont {B.~J.}\ \bibnamefont
  {Min}},\ }\href@noop {} {\  (\bibinfo {year} {2016})},\ \Eprint
  {http://arxiv.org/abs/arXiv:1606.08145} {arXiv:1606.08145} \BibitemShut
  {NoStop}%
\bibitem [{\citenamefont {Cruz-Garza}(2014{\natexlab{b}})}]{cruzgarza2014b}%
  \BibitemOpen
  \bibfield  {author} {\bibinfo {author} {\bibfnamefont {J.~G.}\ \bibnamefont
  {Cruz-Garza}},\ }\href@noop {} {\enquote {\bibinfo {title} {\emph{Forces
  acting on a basketball in flight}},}\ }\bibinfo {howpublished}
  {\url{physicsofbasketball.wordpress.com/2014/05/18/forces-acting-on-a-basketball-in-flight}}
  (\bibinfo {year} {2014}{\natexlab{b}}),\ \bibinfo {note} {online May 18,
  2014}\BibitemShut {NoStop}%
\bibitem [{bri(2014)}]{britannica2014}%
  \BibitemOpen
  \href@noop {} {\enquote {\bibinfo {title} {\emph{Three-point line:
  Basketball. Britannica Online for Kids}},}\ }\bibinfo {howpublished}
  {\url{kids.britannica.com/comptons/art-188763}} (\bibinfo {year} {2014}),\
  \bibinfo {note} {accessed December 16, 2016}\BibitemShut {NoStop}%
\bibitem [{mat(2016)}]{matlab2016}%
  \BibitemOpen
  \href@noop {} {\enquote {\bibinfo {title} {\emph{MATLAB R2016b, The
  MathWorks, Inc., Natick, Massachusetts, United States}},}\ } (\bibinfo {year}
  {2016})\BibitemShut {NoStop}%
\end{thebibliography}%

\end{document}